\documentstyle[11pt,psfig]{article}
\begin{document}
\title{\bf Do Young Neutron Stars Which Show Themselves As AXPs, SGRs and 
Radio Pulsars Accrete?}

\author{S. O. TAGIEVA$\sp1$
\thanks{e-mail:msalima@lan.ab.az}
, E. YAZGAN$\sp2$
\thanks{e-mail:yazgan@astroa.physics.metu.edu.tr}
, and A. ANKAY$\sp2$ \thanks{e-mail:askin@astroa.physics.metu.edu.tr} \\ 
\\{$\sp1$Academy of 
Science, Physics Institute Baku 370143,} \\ {Azerbaijan Republic} \\
{$\sp2$Middle East Technical University, Department of Physics,} \\ {Ankara,
Turkey}}

\date{}
\maketitle

\begin{abstract}
\noindent
We examined the fall-back disk models, and in general accretion,  
proposed to explain the properties 
of anomalous X-ray pulsars (AXPs), soft gamma repeaters (SGRs), and radio 
pulsars (PSRs). We 
checked the possibility of some gas remaining around the neutron star 
after the supernova explosion. We also compared AXPs and SGRs with the X-ray 
pulsars in X-ray binaries. We conclude the existing models of accretion 
from a fall-back disk are insufficient to explain the nature of AXPs/SGRs, 
particularly the SGR bursts. We also discussed the proposed model of 
combination of magnetic dipole radiation and propeller torques in order to 
explain 
the evolution of radio pulsars on the P-\.{P} diagram. We found that the 
predictions of 
this model contradict the observational data. \\ \\ KEY WORDS AXP, SGR, 
PSR, SN 

\end{abstract}

\parindent=0.2in

\newpage

\section{Introduction}
In the last few years soft gamma ray repeaters (SGRs) and anomalous 
X-ray pulsars (AXPs), which are considered as a separate class of X-ray 
pulsars, have attracted much attention (see Mereghetti 2001 and references 
therein). The only important difference between 
SGRs and AXPs is that SGRs have active periods showing gamma ray bursts.
A puzzling property of these objects is that
their X-ray luminosities (L$_x$) are 1-3 orders of magnitude higher
than their rate of rotational energy loss (\.{E}). 
Besides, repeating gamma ray 
bursts of SGRs is the most important property of this class of objects. 
In order to explain the main physical peculiarities of these 
objects, not only a theory overstepping the limits of the standard 
NS/pulsar physics is needed, but also examining another type of NS,   
known as radio quiet NS, with low X-ray luminosity is required. 

There is no radio radiation observed from AXPs/SGRs, only
the upper limits for radio fluxes (and luminosities) are known (Gaensler 
et al. 2001). It does not necessarily mean that these objects do not 
radiate in the radio band only because of failing to observe radio pulses,
because most of the pulsars are born with low luminosities (Lyne et 
al. 1998; Allahverdiev et al. 1997) and also because of the beaming factor.
However, if somehow it is shown that AXPs and SGRs do not radiate in the 
radio band, then there will be a sharp drop in the radio luminosity 
of pulsars including AXPs and SGRs as we go to higher magnetic fields. 
It is difficult to check this possibility because it is not easy to 
increase the number of  such observational data as the number of AXPs 
and SGRs are small and they are located at large distances. 

The so called magnetar model is proposed by Thompson and Duncan (1995) 
as a potential explanation for the nature of AXPs and SGRs. According to 
this model these sources are isolated rotating NSs with surface dipole 
magnetic fields of 10$^{14}$-10$^{15}$ G, extracting their luminosity 
from the decay of the magnetic field. 

The alternative model is that the X-ray luminosity is due to accretion 
from a fall-back disk that is left over from the supernova that formed 
the NS (Chatterjee et al. 2000; Alpar 2001). In this work we will show 
the incompetence of the $existing$ accretion models in explaining the 
AXPs, SGRs, and also radio pulsar properties. In section 2, the 
similarities and differences between SGR bursts and flares of flare stars 
are shown qualitatively and
quantitatively. In section 3, we investigated the possibility of a fossil 
disk remaining around the NS after the SN explosion, and if such a disk 
exists, the effects of this disk on the observed properties of the NS. In 
section 4, we compared the specific angular momentum transfer in AXPs and 
SGRs with the specific angular momentum transfer in X-ray binaries. In 
section 5, after a brief review of history of accretion theory, we showed 
that the proposed propeller mechanism for radio pulsars can not account 
for the evolution of radio pulsars on the P-\.{P} diagram. The $existing$ 
fall-back disk models proposed to explain the nature of AXPs and SGRs are 
not well developed that some predictions of these models contradict  
some of the observational data. These will be shown below. 

\section{Bursts and Flares Due to Magnetic Activities}
In general bursts, and in particular gamma-ray bursts, are common 
astronomical phenomena. In this section we compare SGR bursts with the 
bursts of well known objects in astrophysics.
Bursts are observed, for instance, for the 
Sun, flare stars (UV-Cet type), and T-Tauri stars. 
Flares have fast rise-times of luminosity and longer decay 
times. For all of the flaring objects, the flares  are related to the 
magnetic activities and have short characteristic times ($\sim$minutes 
for UV-Cet stars). Intense flares are followed by less-intensive ones. 
Durations of SGR bursts (10 ms-10 s)
bear resemblance to (taking into 
account the very small size of the NS) the durations of flares;  
roughly, duration of a flare is directly proportional to the size of the 
star. Flares occur on a large portion of the surface of flare stars. 
This must be also true for SGRs since the bursts, particularly the giant 
bursts, can not be produced on a small portion of the neutron star's surface.
Radius of a typical UV-Cet type star is about
R$_{flare}$=10$^{10}$ cm, and of a neutron star is about 
R$_{NS}$=10$^6$ cm. Taking typical flare duration ($\Delta t_{flare}$) to 
be 1 minute, the duration of the SGR burst can be estimated to be 60 ms.
This value is consistent with the observed $\Delta t_{SGR}$ values.
The difference between flares and SGR bursts is the power output, and time 
intervals between the strong flares (for UV Cet stars it is, on the average, 
about 10-100 times shorter compared to SGR bursts). 
The ratio of the energy of outburst to the persistent radiation 
energy for flare stars (UV-Cet type) can be as large as 10$^4$ and for 
SGRs 10$^8$. 
Since AXPs and SGRs are 
considered as one class of objects, we must enquire if AXPs 
also  have outbursts, however, not frequently and/or with low intensity. 

SGR bursts have very large energies in the 
range $\simeq 10^{40-44}$ erg with characteristic timescales of 10 ms-10 
s. Such bursts can not be brought into existence by the $existing$ accretion 
(onto a NS) models and such bursts have never been observed in accreting 
binary 
systems even though X-ray luminosities of some binaries are about two 
orders of magnitude 
higher than the persistent X-ray radiation of AXPs and SGRs. 
It must be noted that, the X-ray bursts in low mass X-ray binaries 
(LMXBs) are due to accretion powered thermonuclear 
reactions on the surfaces of old NSs (Lipunov 1992). For these bursts to 
occur the magnetic field of the burster should be small. The burster 
characteristic  has not
been observed for high mass X-ray binaries (HMXBs). For transient X-ray 
binaries the bursts are due to change in the accretion rate.  
The source of these bursts observed from transients and bursters, which have 
very different physical 
mechanisms, is not the magnetic field of the NS, but accretion. Moreover, 
the spectra of the bursts observed from X-ray binaries and SGRs   
are different.  

As we see above, characteristics of the SGR bursts are similar to the 
characteristics of flares. This is an evidence that the origin of SGR bursts 
is the magnetic field of the NS. If this is the case, 
magnetic field of the NS must be $\sim$10$^{14}$-10$^{15}$ G to power
the SGR bursts with energies of $\sim$10$^{40}$-10$^{44}$ erg (Thompson 
\& Duncan 1995). The $existing$ accretion models avoid to explain the SGR 
bursts and this is the most important handicap of such models.

\section{The Fall-Back Disk Model}
\subsection{Can a Disk Remain Around a Neutron Star After Supernova 
Explosion?} 

The X-ray luminosity (L$_x$) of AXPs and SGRs can be explained by accretion, 
however, 
the compactness of the values of L$_x$ is achieved due to the fact that 
the parameters \.{M} and B are adjustable in a wide range in the 
accretion models. 

In the fall-back disk models, it is very important to know how much mass 
of gas may remain close to the NS after the supernova explosion. 
Gravitational energy of a NS with one M$_\odot$ and a radius about 
1.5$\times10^6$ cm can be estimated as (1-2)$\times10^{53}$ ergs which is 
about 5-10$\%$ of M$_\odot$c$^2$.   
The gravitational energy is converted into heat and rotational energy, 
and  this heating energy, about 10$^{53}$ ergs, transforms into the 
energy of neutrinos and antineutrinos during supernova explosion (Zeldovich 
and Guseinov 1965, Guseinov 1966). 
From the study of SNRs, it is known that 
their kinetic energies lie in the interval $\sim$3$\times 10^{49}-10^{51}$ 
ergs. 
From the observations 
of SNe it is known that the explosion energy, on the average, lies in the 
interval 10$^{50}$-10$^{51}$ ergs and the thrown out mass of gas is
$\sim$(0.5-4)M$_\odot$. 
This shows that only a 
small part of 
the energy of the neutrinos transform into explosion energy, but, in 
principle, this is 
enough to sweep out all the mass which are present near the NS. On the 
other hand, whether some fall-back matter remain near the NS 
(and also its quantity) depends on the velocity distribution of the thrown 
out matter. 

The part of the gravitational energy which is transformed into rotational 
energy of the NS  mainly depends on initial value of angular momentum 
(and its distribution)  
of the collapsed star. Even if we know these values, in order to 
calculate the rotational energy of the NS we must also know 
the value of the magnetic viscosity as well as the dynamics of the 
collapse and the explosion. At this point, there are many 
uncertainties and therefore angular momentum and rotational energy of 
the NS 
should be estimated from the initial period of pulsars. Since, the initial 
period is about 10 ms, initial rotational energy of 
the NS must be about 5$\times10^{51}$ ergs. The collapse transforms the 
angular momentum from the NS (which is about to be born) to external parts 
and this can sweep out the 
surrounding matter (Bisnovatyi-Kogan 1971; Amnuel et al. 1973). 
Actually, observational data do not show presence of gaseous disks or 
planets near single pulsars with a magnetic field in excess of  
10$^{10}$ G. 
Optical observations given in Kaplan et al. (2001) showed that there 
is no optical counterpart of SGR 0526-66 and no accretion disk is 
present around the NS. Moreover, due to precession of the disc around 
the NS and/or inhomogeneities in the disc, variations in the dispersion 
measure values of PSRs must have been observed at least for some of the 
nearby 
young PSRs. Therefore, there is no basis for the presence of any matter 
of fall-back around AXPs and SGRs, but we can not completely exclude the 
possibility of a little bit of gas, which can not be observed, remaining 
near the neutron star.    

\subsection{Are AXPs and SGRs Accreting Systems?}

As we indicated above, an unobservable amount of matter might remain 
around the NS. Below, we examine if such a small amount of matter spins 
up or spins down the NS. 	 
When the collapse begins to slow down, the SN explosion occurs and the shell 
around the NS is thrown away. Due to conservation of angular momentum, the
rotational speed of the shell will decrease as it moves away from the NS.
At the beginning, the magnetic field of the NS is frozen to the shell. 
This slows down the 
NS, and speeds up the shell (Bisnovatyi-Kogan 1971 and Amnuel et al. 
1973). Angular speed of the shell falling back increases. 
Consequently, the NS will be spun-up by the fall-back matter if the 
propeller mechanism is not working simultaneously, i.e. if the fall-back 
matter does not 
lose angular momentum. However, the $existing$ accretion from a fall-back 
disk models do not explain how the fall-back disk loses angular momentum 
instead of gaining it. Models based on fall-back disk should include 
propeller mechanism as well as accretion mechanism working together 
simultaneously. 

In the $existing$ fall-back disk models 
the parameters of the disk  
(related to the magnetic field and rotational speed of NS) is chosen such 
that the X-ray luminosity is about $10^{34-36}$ erg/s and \.{P} 
$\approx10^{-13}-10^{-11}$ s/s.
So, the $existing$ models, in fact, are not related to the properties and 
history of the fall-back matter. 
On the other hand, the only accretion model (onto a 10$^{12}$ G NS) which takes 
into account the small ages of AXPs and SGRs is the model of accretion from 
a fall-back disk.

\section{Comparisons Between AXPs/SGRs and X-ray Pulsars in Binaries}
We examined the differences between
L$_x$/$|$\.{E}$|$ values of AXPs, SGRs, and X-ray pulsars 
in binary systems. Here, L$_x$ is the X-ray luminosity which 
depends only on the rate of accreted mass, \.{M$_x$}, and \.{E} 
is the rate of rotational 
energy change which depends on both \.{M$_x$} and specific angular 
momentum, L$_{spc}$.
Therefore, L$_x$/$|$\.{E}$|$ will be inversely proportional to 
L$_{spc}$. 
In Table 1, L$_x$ and $|$\.{E}$|$ values (and their ratio) of X-ray 
pulsars in high mass X-ray binaries (HMXBs), 
a LMXB, and AXPs/SGRs with known \.{P} values are displayed (we do not 
include transient X-ray binaries in this table, because it is difficult 
to determine the correct value of L$_x$/$|$\.{E}$|$ for such systems). 
For HMXBs, L$_x$/$|$\.{E}$|$ values range from $3\times10^3$ to 
$3\times10^6$, with one exception, namely H0115-737, which has a lower 
L$_x$/$|$\.{E}$|$ value of 65. This is because of the small spin period 
value of this pulsar. There is only 1 LMXB with reliable \.{P} and L$_x$ 
values and its L$_x$/$|$\.{E}$|$ value is (5-7.5)$\times 10^2$, less than 
all of the L$_x$/$|$\.{E}$|$ values of HMXBs which have spin periods 
close to spin periods of AXPs and SGRs.   

Accretion onto a NS can spin it up or down. However, for binary systems 
with the same parameters, we expect that as \.{M$_x$} increases 
\.{P}, $|$\.{E}$|$ and L$_x$ increase. 
There is disk accretion in LMXBs and 
unit mass of the accreted matter has more angular momentum compared 
to the unit mass of the accreted matter in HMXBs since, on 
the average, they have higher orbital velocity. 
Because of this reason, for LMXBs L$_x$/$|$\.{E}$|$ is smaller than for
HMXBs, on the average. Without the propeller effect, 
value of L$_x$/$|$\.{E}$|$ for fall-back matter accreted onto the NS
must not be smaller than for X-ray pulsars in HMXBs if their spin period 
values are close to each other. As seen from Table 1, 4 of 7 AXPs/SGRs 
seem to have lower values than 
LMXBs  that accretion from fall-back matter without propeller effect 
contradicts the observational data. On the other hand, the propeller 
effect must be weak enough not to be observed 
in the plerionic parts of the SNRs, because the SNRs which have
genetic connections with AXPs have pure shell type structures (Gaensler 
et al. 2001; Tagieva \& Ankay 2002).   

\section{The Other Aspects of Accretion: Single Stars}
Classical accretion (onto a single star) theory is a very developed and 
well known subject  
in astrophysics. To explain the radiation from stars, in early 
times, accretion from   
interstellar medium and contraction of stars were proposed. 
Later
it was understood that thermonuclear reactions in the cores of stars are 
the source of radiation of stars, then there was no need for accretion 
anymore. 
On the contrary, it was found that stars have winds, sometimes very 
strong winds.

Salpeter (1964) tried to explain the X-ray sources (when they
were first found by Giacconi et al. 1962) based on accretion 
from interstellar medium onto a single neutron star.
So, accretion onto single stars was called to mind. It is 
well-developed for systems including NSs and black     
holes. 
However, accretion from interstellar medium is found to be wrong, 
because it is found that if there is accretion onto a single NS, then the 
NS becomes hotter and the speed of sound in the surrounding matter 
increases, so that the accretion rate decreases by 6 (Schwartsman 
1970) or 8 (if the magnetic field frozen in the interstellar medium is 
also taken into account, Amnuel and Guseinov 1972) orders of magnitude. 
X-ray sources were explained as accreting binary sytems (Zeldovich and 
Guseinov 1966).

After PSRs were discovered, it was understood that rapidly rotating single 
NSs also produce winds (Pacini 1967). Such winds are much more efficient 
than radiation pressure in sweeping out the surrounding matter around 
PSRs (Lipunov 1992).
Then accretion theory could only work for slowly rotating single NSs with 
low magnetic fields. 
For 30 years, this has been searched for accreting old single NSs in 
optical and soft X-ray bands without considerable success (Danner 1998). 
Also there was no success in finding fluctuating (accreting from 
interstellar medium) single black holes (Schwartsman 1970). Moreover, there 
is no sign of fall-back 
matter in the central parts, particularly in the near-environments, of point 
sources of historical (i.e. very young) SNRs Crab (PSR J0534+2200), 3C58 (RX 
J0201.8+6435, Torii et al. 2000; Bocchino et al. 2001), Cas A (CXO 
J2323+5848, McLaughlin et al. 2000; Kaplan et .al 2001).  

\subsection{Propeller Mechanism for Radio Pulsars}
Despite the facts against accretion onto single NSs given above, 
Alpar et al. (2001) following 
Menou et al. (2001) have proposed that the distribution of radio pulsars 
in P-\.{P} diagram can be explained by the combination of magnetic dipole 
torque and propeller torque of a fall-back disk, instead of pure 
magnetic dipole radiation torque. Alpar et al. (2001) predicted 
the evolutionary tracks (represented with the 2 curves in Figure 1) of 
pulsars on the P-\.{P} diagram. Constant B-field lines are also shown 
in Figure 1 to make a comparison between them. 
According to Peng et al. (1982) and Huang et al. 
(1982) neutrino emission from pulsars and magnetic dipole 
radiation of superfluid neutrons, respectively, also yield  
tracks similar to those of Alpar et al. (2001). 
As seen in Figure 1, down to a minimum period 
derivative value, pulsars follow 
dipole-dominant radiation tracks and after that point 
pure dipole and combined dipole+propeller tracks diverge from each other.
Period derivative becomes \.{P} $\propto 
P^3$ in the propeller-dominant phase. This diversion also reflects 
that, after the dipole-dominant phase is 
over, the ages predicted by the model of Alpar et al. (2001) and the 
characteristic 
ages (which are determined by the effect of pure magnetic dipole 
radiation torque) follow different paths as can be seen in Figure 1. 

The real ages of pulsars are the kinetic ages which are valid for all 
models; the kinetic age is proportional to the distance of the PSR from 
the Galactic plane (see e.g. Lyne \& Graham-Smith 1998). 
So, the dipole+propeller model can be tested by comparing the model's age 
predictions with the kinetic ages. 
According to the dipole+propeller model, older pulsars must be located in 
the upper right part of the P-\.{P} diagram, 
and these pulsars must be far away from the Galactic plane, i.e. their 
kinetic ages must be larger. In order to 
check this, we constructed the P-\.{P} diagram by representing the 
pulsars with $|z|$ $<$ 200 pc with + symbol and the 
pulsars with $|z|$ $>$ 400 pc by open circles in 
Figure 1, where $|z|$ is the distance of pulsar from the Galactic plane. 

In general, the birth places of pulsars are very close to the Galactic 
plane. The average scale height at the time of birth is about 60 pc, 
similar to OB stars' average scale height. However, in the outer parts of 
the Galaxy, the star 
formation regions might deviate from the geometric plane of the Galaxy. 
Optical observations of cepheids with high luminosities and of red 
supergiants located at distances about 5-10 kpc from the Sun, in 
the directions l$\sim200^o-330^o$, showed that the star formation 
regions are located below the Galactic plane by about 300pc, 
and the star formation regions in the directions l$\sim70^o-100^o$ are 
located above the plane by about 400pc.
The star formation regions at about 3-5 kpc from the Sun in the 
directions l$\sim270^o-320^0$ are located about 150 pc below the geometric 
plane of the Galaxy (Berdnikov 1987). These deviations of the locations 
of star formation regions from the Galactic plane have strong
influence on the kinetic ages of young pulsars. 
We did not include the pulsars located in the deviated parts of the star 
formation regions indicated above. We also did not include the pulsars 
beyond 5 kpc due to large uncertainties in the distance measurements.
As seen from Figure 1, along the path (from left to right) of the 
dipole+propeller 
tracks, kinetic ages first increase and then start to decrease. This is 
inconsistent with the propeller model. Therefore, dipole+propeller model 
can not account for the evolution of pulsars on the P-\.{P} 
diagram. Age predictions due to pure magnetic dipole radiation mechanism 
are in 
good agreement with the kinetic ages (see Figure 1). Also, Allakhverdiev 
\& Tagieva (2002) showed that magnetic dipole evolutionary tracks are 
more reliable; pulsars evolve roughly with
constant magnetic field, number of pulsars increase
with characteristic time on constant magnetic field lines.

Only a few of the pulsars with characteristic ages less than $10^6$ years 
are more than 400pc away from the plane (see Figure 1) which could be 
related to their place of birth being high above the Galactic plane (the 
progenitor can be a runaway star) and their speed being large. 

There is also another possibility to test the propeller model of Alpar et 
al. (2001) which predicts the change of \.{P} (\"{P}) to be the 
highest on the propeller-dominant 
parts of the evolutionary tracks. In order to find the braking index 
(n=$\ddot{\Omega}\Omega/\dot{\Omega}^2$, where $\Omega$ is the angular 
frequency of the pulsar) the value of \"{P} must be large. 
About 40 years of observations have shown that, in that part of 
the diagram, \"{P} values of pulsars are not large enough to be measured. 
This also contradicts the propeller model. 

Glitch is a common phenomenon for young radio pulsars. 
According to the model of Alpar et al. (2001) for pulsars with P$>$ 1 s 
and \.{P}$>$ 10$^{-14}$ s/s no glitch must be observed, because 
according to their model these pulsars must be older than about 10$^7$ 
yrs. If glitches from these pulsars are observed, then this will directly 
show that they are young.

\section{Conclusions}
In this work, we have focused on if a residual (fossil) disk may remain 
around single 
NSs (AXPs, SGRs, and radio pulsars) after the supernova explosion and the 
effects of such a disk on the NS properties. We approached the problem 
from various different ways.
We tried to clarify the physical nature of these objects 
and the existing accretion models in the light of observational data. In 
summary, we have showed that: \\
1) The characteristics of SGR bursts are similar to the characteristics of 
flares of flare stars all of which have the same source (origin): the 
magnetic field of the star, but not accretion onto the star. \\
2) After the supernova explosion, only a little amount of mass may remain 
around a single NS. The explosion energy of the thrown out matter is 
enough to sweep out most of the mass around the NS. \\
3) The surrounding matter can also be swept out by angular momentum 
transfer due to magnetic braking and this is supported by observational 
data which do not show gaseous disks or planets (or the effects of these) 
near single pulsars with a magnetic field in excess of 10$^{10}$ Gauss. \\
4) Even if such a little amount of mass remains around the NS, it spins 
up the NS (instead of spinning it down) unless the propeller mechanism is 
working together with accretion simultaneously. We showed this also by 
L$_x$/\.{E} considerations in section 4. 
However, the propeller effect should be weak enough not to be observed 
since all the AXPs genetically connected to SNRs are pure shell 
type remnants. Fall-back 
disk models should be constructed taking these facts into account. \\
5) Comparing the kinetic ages of pulsars with the age predictions of the 
propeller+dipole model for radio pulsars, we showed that there is a 
strong contradiction between them. So, propeller model does not work for 
radio pulsars.

$Acknowledgments$.
It is a pleasure to thank Oktay H. Guseinov for fruitful and stimulating 
discussions. We thank Gennadi S. Bisnovatyi-Kogan for comments on the 
manuscript.
We also thank T\"{U}B\.{I}TAK, the Scientific and Technical Research 
Council of Turkey, for support through TBAG-\c{C}G4.

\clearpage

\begin{table*}
\footnotesize
\centering
\caption[]{The Data of AXPs/SGRs, and X-ray Pulsars in X-ray Binaries with 
Measured \.{P} Values} 
\label{table1}
$$
\begin{array}{p{0.1\linewidth}cccccccc}
\hline
\noalign{\smallskip}
Names & P_{orb} & P & \dot{P} 10^{-11} & L_x 10^{36} & |\dot{E}|=3.94 
10^{78}\frac{|\dot{P}|}{P^3} & L_x/|\dot{E}| 10^3 & Ref. \\
& (d) & (s) & (s/s) & (erg/s) & (erg/s) & &  \\ \hline

H053109-6609.2  & \sim700 & 13.68 & 3.7 
& 2.4  & 5.7 & 5-20 & [1,2,3]  \\
T & & & & (0.1-2.4) & & & \\
LMC & & & & 10 & & & \\
& & & & (2-10) & & & \\ \hline

H0532-664 & 1.41 & 13.5 & 6.1 & 400 &
10 & 400 & [4,5,15,16]  \\
LMC & & & & & & & \\ \hline

L1627-673  & 0.029 & 7.67 & 17 & 2 & 44 & 0.5-0.75 & [8,9,10] \\
Q & & & & (2-10) & & & \\
& & & & 0.033 & & & \\ \hline

H1119-603 & 2.09 & 4.82 & -3.8 & 44 & 140 & 3 & [6,7,20] \\
Q & & & & (2-10) & & & \\ \hline

H0115-737 & 3.89 & 0.71 & -1.6 & 111 & 17000 & 0.065 & [17,18,19] \\
SMC & & & & & & & \\ \hline

H0352 & 580.7 & 836.8 & 420 & 0.006 & 0.0029 & 20 & [11,12,13,14] \\
+309 & & & & (0.1-2.4) & & & \\ \hline

H1538-522 & 3.73 & 529 & 390 & 2.9 & 0.01 & 2900 & [21,22,23,24] \\
& & & & (1-15) & & & \\ \hline

1E1841-045 & & 11.77 & 4.1 & 0.4 & 9.9 & 0.4 & [25,26] \\
AXP & & & & (0.1-12) & & & \\ \hline

J170849-4009 & & 11 & 2.25 & 1 & 6.7 & 1.5 & [27,28] \\
AXP & & & & (0.1-2.4) & & & \\ \hline 

0142+614 & & 8.69 & \sim0.22 & 0.1 & 1.2 & 0.8 & [41,42,43] \\
AXP & & & & (0.1-2.4) & & & \\ \hline

2259+587 & & 6.98 & 0.06 & 0.2 & 0.69 & 3 & [29,30,31,32] \\
AXP & & & & (0.5-4) & & & \\ \hline

1048.1-5937 & & 6.45 & \sim2 & 0.0063-0.3 & 29 & 0.002-0.1 & 
[33,34,35,36] \\ 
AXP & & & & (0.1-2.4) & & & \\ \hline

1806-20 & & 7.47 & 8.3 & 1 & 78 & 0.13 & [37] \\
SGR & & & & (0.5-10) & & & \\ \hline

1900+14 & & 5.16 & 6 & 0.1 & 320 & 0.003 & [38,39,40] \\
SGR & & & & (2-10) & & & \\ 
\hline
            \noalign{\smallskip}
         \end{array}
     $$
\end{table*}
\clearpage
\newpage
\begin{list}{}{}
\item[$^{\mathrm{}}$]

[1] Haberl et al. 1995; [2] Hanson et al. 1989; [3] Burderi et al. 1998; 
[4] Levine et al. 1991; [5] Woo et al. 1996; [6] Burderi et al. 2000; [7] 
Tsunemi et al. 1996; [8] Mereghetti \& Stella 1995; [9] Chakrabarty et 
al. 1997; [10] Angelini et al. 1995; [11] Hutchings et al. 1974; [12] 
Mavromatakis 1993; [13] Robba et al. 1996; [14] Weisskopf 1984; [15] Li 
et al. 1978; [16] Vrtilek et al. 1997; [17] Tjemkes et al. 1986; [18] 
Yokogawa et al. 2000; [19] Bonnet-Bidaut \& van der Klis 1981; [20] 
Kelley et al. 1983; [21] Clark 2000; [22] Clark et al. 1994; [23] Rubin 
et al. 1997; [24] Robba et al. 1992; 
[25] Vasisht \& Gotthelf 1997; [26] Gotthelf et al. 1999; [27] Sugizaki 
et al. 1997; [28] Israel et al. 1999a; [29] Baykal \& Swank 1996; [30] 
Fahlman \& Gregory 1981; [31] Kaspi et al. 1999; [32] Morini et al. 1988; 
[33] Corbet \& Mihara 1997; [34] Seward, et al. 1986; [35] Mereghetti, 
1995; [36] Oosterbroek et al. 1998; [37] Kouveliotou et al. 1998; [38] 
Kouveliotou et al. 1999; [39] Sonobe et al. 1994; [40] Marsden et al. 
1999; [41] Israel et al. 1999b; [42] White et al. 1996; [43] Israel et 
al. 1994. 

\end{list}

\clearpage
\clearpage

\begin{figure}
\centerline{\psfig{file=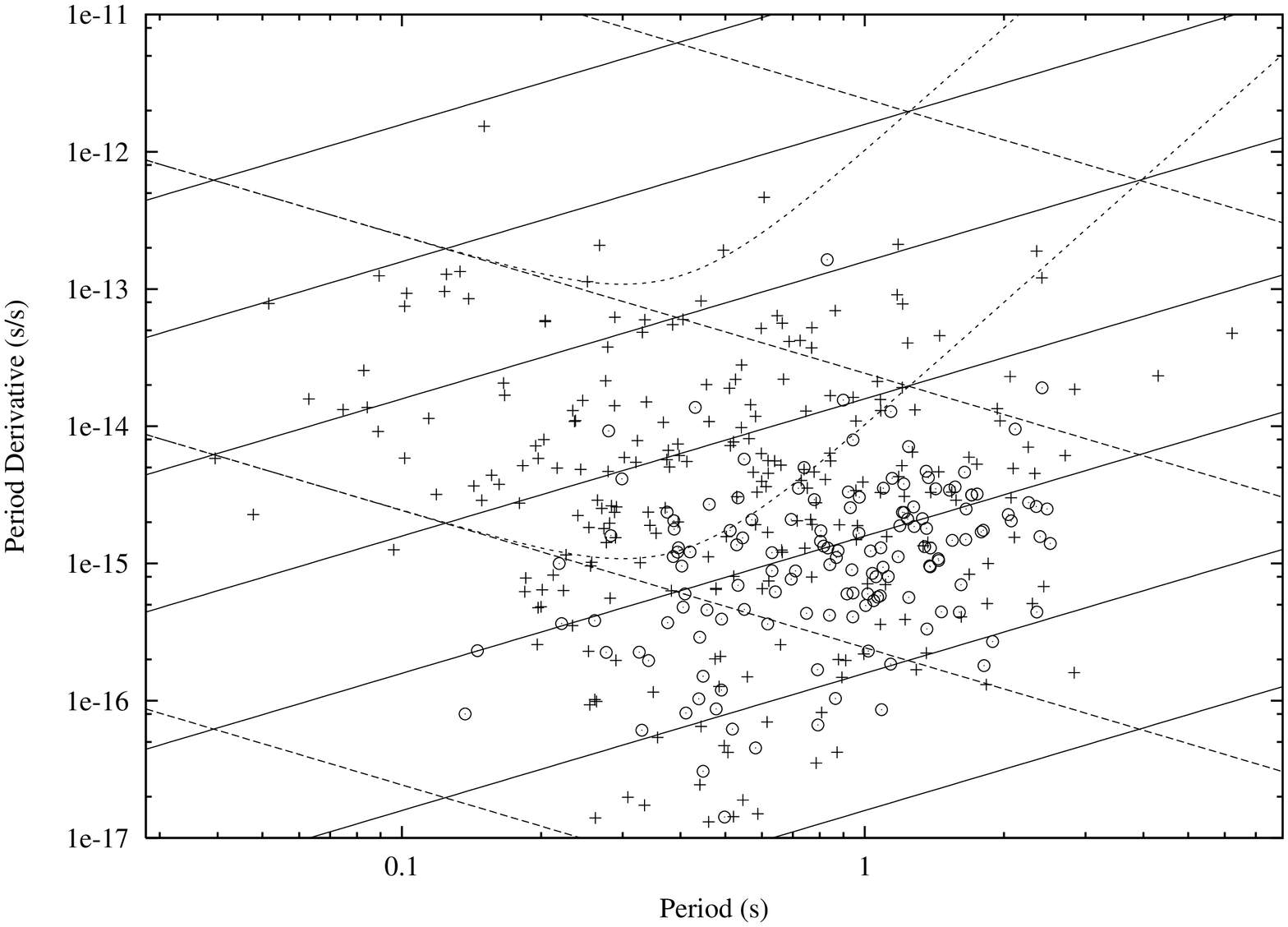,width=18cm,angle=-90}}
{Figure 1. P-\.{P} diagram for PSRs up to 4 kpc.  
Characteristic age lines (calculated from pure magnetic dipole radiation) 
are from 10$^3$ to 10$^9$ yrs. Constant magnetic dipole 
field lines range from 5$\times 10^{10}$-5$\times 10^{13}$ gauss. The two 
curves, taken from Alpar et al. (2001), represent the  
evolution of pulsars due to dipole+propeller torques.}  
\end{figure}

\end{document}